\documentclass[fleqn,10pt]{wlscirep}
\usepackage[utf8]{inputenc}
\usepackage[T1]{fontenc}
\usepackage{cite}
\usepackage{amsmath,amssymb,amsfonts}
\usepackage{cuted}
\usepackage{physics}
\usepackage{algorithmic}
\usepackage{graphicx}
\usepackage{textcomp}
\usepackage{comment}
\usepackage{float}

\title{Real-Time Polarization Control for Satellite QKD with Liquid-Crystal Beacon Stabilization}

\author[1,*]{Ondrej Klicnik}
\author[2]{Alessandro Zannotti}
\author[2]{Yannick Folwill}
\author[2]{Oliver de Vries}
\author[1]{Petr Munster}
\author[1]{Tomas Horvath}
\affil[1]{Brno University of Technology, Department of Telecommunications, Brno, 616 00, Czech Republic}
\affil[2]{Quantum Optics Jena, Department of Research and Development, Jena, 07745, Germany}

\affil[*]{ondrej.klicnik@vut.cz}

\keywords{Satellite QKD, Entanglement-based QKD, BB84, Polarization compensation, Polarimetry, Trusted node}

\begin{abstract}
Polarization instability is a critical challenge for polarization-entangled satellite quantum key distribution (QKD), where atmospheric effects and platform motion continuously distort photon polarization. To maintain entanglement fidelity, these transformations must be accurately identified and compensated prior to detection. In this work, a compact and fast polarization-compensation approach based on liquid-crystal (LC) variable retarders is presented, using a co-propagating classical reference signal (beacon) for real-time polarization tracking. An LC-based polarimeter is implemented, and its performance is evaluated using both direct and Fourier-based Stokes parameter reconstruction. Experimental results indicate that accurate polarization estimation can be achieved with a limited number of measurements, enabling a favorable trade-off between speed and precision. The impact of liquid-crystal switching dynamics is also analyzed, highlighting the importance of selecting appropriate operating conditions for real-time applications. In addition, the effect of polarimetric inaccuracies on QKD performance is assessed through simulations of an entanglement-based protocol. The results show that only a moderate increase in quantum-bit error rate is introduced, while remaining compatible with secure key distribution. These findings demonstrate that LC-based polarization control represents an efficient and practical solution for real-time compensation in satellite QKD systems.

\end{abstract}
\begin{document}

\flushbottom
\maketitle
%
%
\thispagestyle{empty}


\section*{Introduction}

Quantum key distribution (QKD) has matured from laboratory demonstrations to pilots of metropolitan fiber networks. However, its operational range is limited by channel losses, photon source design, and detector capabilities. To extend loss-driven boundaries and enable coverage that spans nations, multiple space-based QKD missions have been launched since 2016. Europe has announced the \emph{EuroQCI} program, which will fuse terrestrial fiber networks with a dedicated satellite layer, starting with the prototype mission \emph{Eagle-1}, planned for launch in early 2026~\cite{EuroQCI2025,Eagle1ESA2024}.

Satellites currently implement either prepare-and-measure BB84/B92 or entanglement-based BBM92 protocols to distribute quantum keys to ground networks~\cite{Liao2018, Lu2022, Zhang2024, oneillSEAQUEArrivesSpace2024}. In prepare-and-measure architectures, keys are independently established between the satellite and each ground station via a single down-link from low-Earth orbit (LEO, ~300–500 km). The single-pass channel loss enables significantly higher secret-key rates and global coverage~\cite{Liao2018, Lu2022}. However, the satellite functions as a trusted node, representing a potential vulnerability, as space assets may be compromised or accessed remotely~\cite{Salvail2010}.

Entanglement-based links, by contrast, distribute photon pairs to two ground stations simultaneously. Although the double down-link introduces higher channel loss and leads to lower key rates, this architecture avoids the need to trust the satellite, enabling true end-to-end security. Current demonstrations are limited to~1120 km due to joint line-of-sight constraints and photon-loss scaling~\cite{Liao2017}.

On Earth, terrestrial fiber-based QKD forms the backbone of emerging quantum communication networks. Fiber links provide stable, scalable urban-scale connectivity and continuous operation independent of satellite passes. While fundamental attenuation limits restrict individual fiber spans to typically a few hundred kilometers~\cite{c1}, advanced protocols such as twin-field QKD have extended secure transmission to 830~km without trusted nodes~\cite{c3}. Rather than replacing fiber infrastructure, satellite QKD complements it by enabling inter-continental and remote-region coverage beyond the reach of terrestrial links.

Satellite QKD therefore plays a critical role in extending the geographical reach of quantum-secured networks~\cite{Zhang2024}. As an example, the Micius mission integrates both BB84 and entangled-photon sources, demonstrating 1120 km entanglement-based QKD without a trusted node~\cite{Lu2022,d3,Liao2018}. In addition, operating as a BB84 transmitter and trusted relay, Micius enabled secure key exchange between Beijing and Vienna over ~7500 km~\cite{d5}, highlighting how fiber-based and satellite-based QKD infrastructures together form the foundation of a globally connected quantum-communication network. Despite enabling long-distance secure key distribution, satellite QKD faces practical challenges such as precise timing synchronization, channel-induced polarization fluctuations, and dynamic link stabilization, motivating the need for effective polarization-compensation strategies. In a~satellite downlink, polarization instability arises because the photons propagate through a time-varying optical system. Satellite motion, changing transmitter–receiver geometry, and polarization effects in the propagation channel and receiver optics introduce time-dependent transformations that accumulate along the optical path. As a result, the received polarization state can deviate from the intended measurement basis, leading to reduced correlation visibility and an increased quantum-bit error rate.


Contemporary satellite-based QKD systems follow two principal strategies for mitigating the polarization drifts introduced by telescopes and the turbulent atmosphere.  
The first, calibration tomography, sends a short burst of bright reference pulses in four polarization states before every key-exchange frame. From the measured Stokes vectors the receiver reconstructs the Mueller matrix of the channel and immediately applies its inverse with local optics.  
This technique underpins both the Canadian \textit{QEYSSat} mission (120 kg micro-satellite) as described in~\cite{QEYSSat} and the German \textit{QUBE} mission (3U CubeSat with less than 3.5~kg)~described here~\cite{qube1,qube2,qube3,qube4}.

An alternative approach to polarization stabilization employs a continuous beacon laser — a bright classical beam co-propagating with the quantum photons. Its instantaneous polarization state is monitored in real time, and compensating elements are dynamically adjusted to correct the observed rotation. This method has been demonstrated on the Micius satellite (630 kg), which used a motorized wave-plate stack to actively cancel polarization drift~\cite{Micius}. Similarly, nanosatellite experiments such as the Singaporean \textit{SpooQy-1} (3U CubeSat, 2.6 kg) have explored liquid-crystal-based (LC) polarization control, though limited to internal loops without down-link correction~\cite{Spooqy1,Spooqy2}. LC-based polarization compensation devices capable of implementing arbitrary polarization transformations have also been proposed for satellite QKD systems~\cite{JimenezGirela2025}, highlighting the potential of fully electronic, compact solutions. For clarity, the main polarization-stabilization techniques relevant to satellite QKD are briefly summarized in Table~\ref{tab:polcomp_methods}.

\begin{table}[H]
\centering
\caption{Main polarization-stabilization techniques relevant to satellite QKD.}
\begin{tabular}{p{0.3\linewidth} p{0.62\linewidth}}
\hline
\textbf{Technique} & \textbf{Brief description} \\
\hline
Pre-frame tomography &
Polarization drift is estimated from reference pulses transmitted before each key-exchange frame. \\

Beacon-based mechanical control &
Polarization drift is estimated from a co-propagating classical beacon and corrected using rotating wave plates. \\

Beacon-based LC control &
Polarization drift is estimated from a co-propagating classical beacon and corrected using liquid-crystal retarders. \\
\hline
\end{tabular}
\label{tab:polcomp_methods}
\end{table}

Against this background, the \textit{CubEniK} project advances beacon-based polarization stabilization by replacing moving optics with two low-voltage liquid-crystal variable retarders, creating a simpler, fully electronic system. While a similar LC-based compensation approach is employed, the main focus of this work is on polarization monitoring, realized through a beacon-based polarimetric scheme. By leveraging the same liquid-crystal technology used for beacon-laser polarimetry, \textit{CubEniK} enables real-time, in-flight down-link correction and allows polarization-control settings to be directly transferred and adapted for the quantum channel.

Further miniaturization of polarization-control subsystems may be achieved in the future using emerging flat-optics approaches. In particular, metasurface-based devices, composed of subwavelength nanostructures~\cite{Neshev2018}, have demonstrated highly efficient and flexible manipulation of optical wavefronts. These structures enable control over multiple degrees of freedom of light, including polarization~\cite{Yuan2024} and spin–orbit interactions, i.e., coupling between polarization and spatial properties of light~\cite{Yuan2025SpinOrbit}. While these approaches are currently predominantly static, they represent a promising direction for future compact and integrated implementations of polarization-control systems.

\clearpage
\section{Polarization monitoring and compensation }
In polarization-encoded QKD, any uncontrolled change of the polarization state acts as basis misalignment. In satellite links, these changes arise from satellite motion, varying propagation geometry, and channel effects, and therefore the polarization state must be continuously monitored and compensated. This requirement holds for both fiber-based and free-space links, although the practical implementation differs between them. Polarization compensation protocols can be systematically divided into two phases.

\begin{itemize}
    \item \textbf{Monitoring} – The first task is to identify the change in polarization, which could be done using a polarization-sensitive measurement or a specific polarimetric method to obtain the exact polarization state. These methods differ in speed, accuracy, and the physical size and weight of the required equipment. 
    \item \textbf{Compensation} – If the desired polarization state is known, the observed deviation can be used to actively adjust the polarization through appropriate calculations or control algorithms. Such adjustments are typically implemented using polarization controllers. As a result, the system is generally brought closer to the original or desired polarization state.
\end{itemize}
Given the inherently low power of quantum signals, conventional approaches, such as diverting part of the optical power to a polarimeter for real-time analysis, are not applicable. Depending on whether a fiber-based or free-space system is used, different compensation strategies must be employed.

\section{CubEniK setup}
The \textit{CubEniK} project considers an entanglement-based satellite QKD architecture in which an entangled-photon source (EPS) and a classical reference beacon are placed on board the satellite, while two optical ground stations perform polarization stabilization and quantum-state analysis. The beacon co-propagates with the quantum channel and is used to estimate channel-induced polarization drift in real time. Figure~\ref{fig:CompleteSystem} shows the receiver-side architecture of one ground station. It consists of a~polarization compensation module for beacon-based monitoring and correction, and a polarization analysis module for the subsequent measurement of the quantum states. The main focus of the present paper is the polarization compensation module, in particular the LC-based polarimeter used for polarization monitoring, while the polarization analysis module is included to show how the compensated quantum signal interfaces with the QKD receiver. Both ground stations are assumed to have the same architecture and are commonly denoted Alice and Bob. Each station consists of two main subsystems:

\begin{itemize}
    \item \textbf{Polarization Compensation Module (PCM)} – This subsystem enables real-time polarization monitoring and correction. It consists of an LC-based polarimeter that measures the instantaneous polarization state of the reference beam and estimates its deviation from the desired state. This information can then be used by a polarization compensator to determine the correction applied to the quantum channel. The exact relation between the measured polarization state and the compensator settings is not universal, since it depends on the specific compensator design, the optical layout, and whether the beacon and quantum channels operate at the same wavelength. In addition, the PCM includes an optical separation unit that isolates the reference signal from the quantum channel. This can be implemented either passively using wavelength-division multiplexing or actively using time-division multiplexing, ensuring that both signals share the same path while remaining distinguishable at the receiver. As shown in Figure~\ref{fig:CompleteSystem}, the LC-based polarimeter consists of a~beam splitter (BS$_1$) that diverts a fraction of the incoming signal to the optical power meter (OPM$_1$), which monitors intensity fluctuations. The transmitted part of the signal then passes through two liquid-crystal retarders (LC$_1$ and LC$_2$) used for polarization analysis, followed by a polarizing beam splitter (PBS$_1$), which transmits the horizontal polarization component while filtering out the vertical one. The transmitted intensity is then detected by the second optical power meter (OPM$_2$). The LC-based polarization compensator is formed by three electrically controlled retarders (LC$_3$--LC$_5$) used for polarization correction.

\item \textbf{Polarization Analysis Module (PAM)} – This is a standard E91 receiver module that performs polarization-resolved detection of the entangled photons, typically using two orthogonal bases. As shown in Figure~\ref{fig:CompleteSystem}, the PAM contains a beam splitter (BS$_2$) for basis routing, a half-wave plate (HWP) for basis transformation, two polarizing beam splitters (PBS$_2$ and PBS$_3$) separating orthogonal polarization components in the two analysis arms, four single-photon avalanche diodes (SPADs) for photon detection, and the time-tagging module (TTM), which records the detection times for coincidence evaluation. The four detectors correspond to projections onto the $\lvert H\rangle$, $\lvert V\rangle$, $\lvert D\rangle$, and $\lvert A\rangle$ polarization states.
\end{itemize}

\begin{figure}[H]
    \centering
    \includegraphics[width=1\linewidth]{PC_free_space_pol_compensation_labels.pdf}
    \caption{Both ground station modules: PCM (including polarimeter) and~PAM. }
    \label{fig:CompleteSystem}
\end{figure}

\section{Stokes and Mueller calculus}
Among various mathematical and graphical representations, the polarization state of light is commonly described by the Stokes vector $\mathbf{S}$, while the effect of optical elements on polarization is characterized by a corresponding Mueller matrix $\mathbf{M}$. Graphically, Stokes vectors can be visualized on the surface of the so-called Poincaré sphere. The resulting polarization $\mathbf{S_\text{out}}$ of light after passing through the optical component can then be calculated as the matrix $\mathbf{M}$ multiplied by the vector $\mathbf{S_\text{in}}$ from the right~\cite{Aiello2006}.
\begin{equation}
\mathbf{S_\text{out}}=\mathbf{M} \cdot \mathbf{S_\text{in}}
\label{eq1}
\end{equation}

Here, 
$\mathbf{M}$
can be understood as an effective time-dependent transformation of the full optical path, including the transmitter optics, propagation channel, and receiver optics, where polarization instability then corresponds to temporal variation of this transformation.
The Stokes vector itself consists of four components, also referred to as Stokes parameters. While the initial component quantifies the total intensity of the specified light source, the subsequent elements delineate a particular form of polarization.

\begin{equation}
\mathbf{S}=\begin{pmatrix}
S_0\\
S_1\\
S_2\\
S_3\\
\end{pmatrix}
\label{eq2}
\end{equation}
Here, $S_0$ denotes the total intensity, while $S_1$, $S_2$, and $S_3$ describe the differences between orthogonal polarization components: horizontal versus vertical, diagonal versus anti-diagonal, and right- versus left-circular, respectively. As the Stokes vector is usually  normalized, it holds that the total intensity $S_0=1$. The other components must be less than or equal to one, as the following equation applies~\cite{Aiello2006}. 
\begin{equation}
S_0^2 \ge S_1^2 + S_2^2 + S_3^2
\label{eq3}
\end{equation}

These parameters can additionally be visualized through the Poincaré sphere representation, which finds a parallel in the Bloch sphere, widely adopted in the field of quantum mechanics. This model offers a three-dimensional visualization of polarization, where the partial Stokes parameters are represented along the three orthogonal axes, with the parameter $S_0$ determining the sphere's radius. In instances where the Stokes vector undergoes normalization, the sphere is given a~radius of one, thus establishing a diameter of two. 

\section{Types of polarimetry}
Standard polarimetric methods can be classified as division-of-time polarimeters that usually require the use of rotating components, such as a quarter-wave plate (QWP) in~\cite{Schaefer2007, Dong2020}, to determine the Stokes vector. However, the use of such components is not suitable in practice because of their size and especially speed (response time). Another thing to consider are static division-of-space polarimeters~\cite{Dong2020} with multiple detectors. These can achieve significantly faster measurements because the components do not have to be reconfigured. Thus, multiple Stokes parameters can be measured at once. However, their disadvantage is the complexity of the components and mutual fine-tuning, which adds greater uncertainty to the measurement. In addition, the expected low-power signal would have to be split across more channels, which would result in lower sensitivity.

A~solution may therefore be the use of liquid crystals, whose effect on polarization does not depend on the rotation of the component but on the applied voltage. Since LCs are already used in the case of a compensator (already built and tested), their use here also has a positive effect on the overall complexity of the system. However, for LCs to work properly, it is first necessary to accurately characterize the mutual relation between retardance ($\delta$) and voltage ($V$) as described in \cite{Schnoor2020}. In this way, it is possible to consider the retardance as a~function of voltage. Even here, however, the accuracy of polarimetric measurements depends on its duration, especially the switching time of the LCs. If the measurement time is too short, the influence of the voltage may not be sufficiently present and the LC may not be properly tuned.

In addition to the physical connection and selection of components, it is also necessary to select the correct analytical method. These may require different amounts of individual measurements and ultimately have a large impact on accuracy.

\section{Proof-of-principle}
For practical use, a test setup similar to the rotating quarter-wave plate method presented here \cite{Schaefer2007} was designed. As can be seen in Figure~\ref{fig:PC_free_space_test_setup}, the experimental setup consists of a~polarization state generator (PSG) and a LC-based polarimeter (PLC). The PSG is used to generate an arbitrary polarization state and incorporates an 810 nm laser. The laser beam is first collimated and then propagated through free space via several optical components. The first of these is a polarizing beam splitter PBS$_0$, which is used to filter out the vertical polarization component. Subsequently, a quarter-wave plate ($\delta=\tfrac{\pi}{2}$) and a~half-wave plate ($\delta=\pi$) are employed, both mounted in motorized rotation stages. The orientation of these waveplates can be controlled from a computer via Python scripts.

\begin{figure}[h]
    \centering
    \includegraphics[width=1\linewidth]{PC_free_space_test_setup_labels.pdf}
    \caption{Polarimetric setup. On the left is a source that is capable of generating any polarization state. On the right is a~polarimeter with two liquid crystals rotated 45° to each other.}
    \label{fig:PC_free_space_test_setup}
\end{figure}

On the right side of the schematic, the same polarimeter as shown previously in Figure~\ref{fig:CompleteSystem} is located. Its purpose is to determine the input polarization state based on intensity variations under different configurations. The light first passes through a~beam splitter BS$_1$, where a small portion of the optical power is reflected towards a~power meter OPM$_1$ to monitor power stability.

This is followed by two liquid crystal retarders LC$_1$ and LC$_2$, with the first aligned at \(0^\circ\) and the second at \(45^\circ\). This configuration defines two measurement bases that allow the entire polarization state space to be resolved. For example, if a~horizontally polarized beam enters the first LC retarder (oriented at \(0^\circ\)), the polarization lies entirely along its fast axis. As a~result, no relative phase delay (retardance \(\delta\)) is introduced between the orthogonal components, and only a global phase shift occurs. The polarization of light thus remains unchanged. The retardance of each LC element is controlled by a~dedicated external controller which is connected to the LC retarders and a PC. The controller is operated via Python scripts executed on the PC. At the output, a polarizing beam splitter PBS$_1$ acts as an analyzer, followed by a power meter OPM$_2$ for detection. 

To ensure correct operation of the polarimeter, not only the hardware is required, but also a derived set of equations that allow the polarization state to be reconstructed. The behavior of the polarimeter and the input–output polarization transformation are modeled using the Stokes–Mueller formalism as:
\begin{equation}
\mathbf{S}_\text{out}=\mathbf{M_\text{PBS1}} \cdot \mathbf{M_\text{LC2}} \cdot \mathbf{M_\text{LC1}} \cdot \mathbf{S}_\text{in}
\label{}
\end{equation}

Since it is further known that the intensity (at the final power meter) can be expressed from the Stokes vector as $I = [1,\,0,\,0,\,0] \cdot \mathbf{S}_\text{out}$, by applying the Stokes–Mueller calculus the equation~\ref{eq4} can be derived:
\begin{equation}
\begin{split}
I(\delta_1,\delta_2)
&= \tfrac{1}{2} \Big(
    S_0 + S_1\cos\delta_2 + S_2\sin\delta_2\sin\delta_1 + S_3\sin\delta_2\cos\delta_1\Big)
\end{split}
\label{eq4}
\end{equation}

Here, $\delta_1$ indicates a fixed retardation value for a horizontally positioned LC and $\delta_2$ for a 45° rotated LC. Since liquid crystals are often supplied as a sandwich consisting of three LCs at 0°, 45° and 0° rotation, it may be convenient to use such a~configuration. For this case, Equation \ref{eq4} holds when the retardance of the last LC $\delta_3$ is set to 0°. Based on this equation, two polarimetric analysis methods can be derived.  

\subsection{Direct analysis}
In order to determine all four Stokes parameters during the polarimetric measurement, the intensity equation must be evaluated under four distinct settings of the LC retarders. This yields four independent intensity measurements that are used to construct a~system of equations. From this system, the Stokes parameters can be extracted. The parameter \(S_0\) appears in all expressions, as it does not depend on the retardance values \(\delta_1\) and \(\delta_2\). 
\begin{equation}
\begin{split} 
I_0 &= I \left( \delta_1 = arb, \delta_2 = 0 \right) = \tfrac{1}{2}(S_0+S_1)  \\
I_1 &= I \left( \delta_1 = arb, \delta_2 = \pi \right) = \tfrac{1}{2}(S_0-S_1)\\
I_2 &= I \left( \delta_1 = \tfrac{\pi}{2}, \delta_2 = \tfrac{\pi}{2} \right) = \tfrac{1}{2}(S_0+S_2) \\
I_3 &= I \left( \delta_1 = 0, \delta_2 = \tfrac{\pi}{2}\right) = \tfrac{1}{2}(S_0+S_3)
\end{split}
\label{eq5}
\end{equation}

Now, the Stokes parameters can be extracted as in Equation~\ref{eq6} below. Finally, the parameters must be normalized with respect to \(S_0\).
\begin{equation}
\begin{split} 
S_0 &= I_0 + I_1  \\
S_1 &= I_0 - I_1 \\
S_2 &= 2 I_2 - S_0 \\
S_3 &= 2 I_3 - S_0
\end{split}
\label{eq6}
\end{equation}

In this case, only four measurements are used. However, in theory, the accuracy of the reconstructed Stokes parameters can be improved by increasing the number of measurements. This is because additional data points provide redundancy, which helps to suppress the effects of noise and measurement uncertainties.

\clearpage
\subsection{Fourier analysis}
As an alternative approach, Fourier analysis can be employed. Specifically, the derived intensity expression can be reformulated as a trigonometric Fourier series \(f(t)\), representing a periodic signal~\cite{aaa}.
\begin{equation}
f(t) = a_0 + \sum_{n=1}^{\infty} \left[ a_n \cos(n \omega t) + b_n \sin(n \omega t) \right]
\label{eq:Fourier_series}
\end{equation}

In this function, the Fourier coefficients can generally be expressed using the integral form as in Equation \ref{eq:Fourier_coefficients}. Here, \(T\) denotes the period of the function.
\begin{equation}
\begin{split} 
a_0 &= \frac{1}{T} \int_{0}^{T} f(t)\, dt \\
a_n &= \frac{2}{T} \int_{0}^{T} f(t)\, \cos( \frac{2\pi n t}{T})\, dt \\
b_n &= \frac{2}{T} \int_{0}^{T} f(t)\, \sin( \frac{2\pi n t}{T})\, dt
\end{split}
\label{eq:Fourier_coefficients}
\end{equation}

To bring the expression closer to the form of Equation~\ref{eq:Fourier_series}, first the case where \(\delta_1 = \delta_2\) is considered. This assumption simplifies Equation~\ref{eq4} to the form shown in Equation~\ref{eq:intensity_FS}. 
\begin{equation}
I(\delta) = \tfrac{S_0}{2} + \tfrac{S_2}{4} + \tfrac{S_1}{2}\cos\delta - \tfrac{S_2}{4}\cos2\delta + \tfrac{S_3}{4}\sin2\delta
\label{eq:intensity_FS}
\end{equation}

If the resulting terms are then expressed as Fourier coefficients: $a_0=\frac{S_0}{2}+\frac{S_2}{4}$, $a_1=\frac{S_1}{2}$, $a_2=\frac{-S_2}{4}$ and $b_2=\frac{S_3}{4}$, the expression can be rewritten as shown in Equation~\ref{eq:intensity_FS_FC}.
\begin{equation}
I(\delta)=a_0+a_1\cos{\delta} + a_2 \cos{2\delta} + b_2 \sin{2\delta}
\label{eq:intensity_FS_FC}
\end{equation}

The objective of the polarimetry in this section is to acquire a set of measurements for different values of \(\delta\).
The integral appearing in the Fourier coefficients in Equation~\ref{eq:Fourier_coefficients} is therefore replaced by a Riemann sum.
Consequently, a complete continuous function is not obtained, instead, a finite set of samples is produced, and the signal is thus discrete. For this reason the sampling step is defined as \(\Delta\delta = \tfrac{2\pi}{N}\) with sample positions \(\delta_i = \tfrac{2\pi i}{N}\), \(i=0,\dots,N-1\),
where \(N\) denotes the number of samples (i.e., measurements).
Noting that both \(\sin\) and \(\cos\) are \(2\pi\)-periodic, and using Equation~\ref{eq:intensity_FS_FC}
together with the definitions in Equation~\ref{eq:Fourier_coefficients},
the Fourier coefficients can be obtained from the discrete measurements \(I(\delta_i)\) as:
\begin{equation}
\begin{split} 
a_0 &= \frac{1}{2\pi}\int^{2\pi}_{0} I(\delta)\,  d\delta \approx \frac{1}{2\pi} \sum_{i=0}^{N-1} I(\delta_i)  \Delta\delta  \\
a_1 &= \frac{1}{\pi}\int^{2\pi}_{0} I(\delta)\cos{\delta}\, d\delta \approx \frac{1}{\pi} \sum_{i=0}^{N-1} I(\delta_i) \cos{\delta_i}  \Delta\delta \\
a_2 &= \frac{1}{\pi}\int^{2\pi}_{0} I(\delta)\cos{2\delta}\, d\delta \approx \frac{1}{\pi} \sum_{i=0}^{N-1} I(\delta_i)  \cos{2\delta_i}  \Delta\delta \\
b_2 &= \frac{1}{\pi}\int^{2\pi}_{0} I(\delta)\sin{2\delta}\, d\delta \approx \frac{1}{\pi} \sum_{i=0}^{N-1} I(\delta_i) \sin{2\delta_i} \Delta\delta
\end{split}
\label{}
\end{equation}

The Stokes parameters of the measured light can then be readily determined for any number of samples or measurements as below. In this case as well, the Stokes vector must be normalized with respect to \(S_0\).
 \begin{equation}
\begin{split} 
S_0 &= 2a_0 + 2a_2  \\
S_1 &= 2a_1 \\
S_2 &= -4a_2\\
S_3 &= 4b_2
\end{split}
\label{}
\end{equation}

\section{Accuracy comparison}
For the proper functioning of any practical system, an appropriate compromise between measurement accuracy and measurement time must be found. The duration of the measurements is then determined mainly by the number of subsidiary intensity measurements and subsequently by the switching time of the LCs. In this case a~total of 4, 8, 16, and 32 measurements were performed for each sent polarization state. The first set was evaluated by means of direct analysis, whereas the remaining sets were evaluated using Fourier analysis. For this reason, two separate automated measurements were designed. In both cases, however, they consist of: 

\begin{enumerate}
    \item \textbf{Calculating Stokes vector} corresponding to the expected PSG output.
    \item \textbf{Measuring Stokes vector} of polarized light, using direct analysis for 4 values and Fourier analysis for 8, 16, and 32 values.
    \item \textbf{Calculating norm} between the two Stokes vectors.
\end{enumerate}
The output is thus a norm interpreted as a distance metric, indicating how close the measured result is to the theoretical calculation.

\subsection{Calculating Stokes vector}
Figure~\ref{fig:PC_free_space_test_setup} shows that the calculated light entering the polarimeter, denoted $\mathbf{S_\text{C}}$, originates as horizontally polarized light $\mathbf{S_\text{H}}$ coming from the PBS$_0$ in the PSG. Subsequently, this light is transformed to the desired ellipticity using the QWP, represented by matrix $\mathbf{M_\text{QWP}}$ and further rotated using the HWP matrix $\mathbf{M_\text{HWP}}$. The resulting state $\mathbf{S_\text{C}}$ is then calculated using matrix multiplication below:
\begin{equation}
\mathbf{S_\text{C}}=\mathbf{M_\text{HWP}} \cdot \mathbf{M_\text{QWP}} \cdot \mathbf{S_\text{H}}
\label{}
\end{equation}

\subsection{Measuring Stokes vector}
In practice, however, the measured result may differ from the calculated one, which may be mainly due to the desire to save time, either due to a shorter time for switching LCs or a smaller number of partial measurements. These polarimetric measurements correspond to the direct and Fourier analysis.

\subsection{Calculating norm}
In order to see how accurate the measurement is, it is necessary to compare the Stokes vector resulting from the measurement with the calculated vector. To do this, the difference between the measured and the calculated vector must first be obtained, denoted as $\mathbf{S_\text{D}}$, and defined as:
\begin{equation}
\mathbf{S_\text{D}} = \mathbf{S_\text{C}} - \mathbf{S_\text{M}} =
\begin{bmatrix}
S_{0C}\\
S_{1C}\\
S_{2C}\\
S_{3C}
\end{bmatrix}
-
\begin{bmatrix}
S_{0M}\\
S_{1M}\\
S_{2M}\\
S_{3M}
\end{bmatrix}
=
\begin{bmatrix}
S_{0C} - S_{0M}\\
S_{1C} - S_{1M}\\
S_{2C} - S_{2M}\\
S_{3C} - S_{3M}
\end{bmatrix}
\label{}
\end{equation}

From this difference, the Euclidean norm \( \lVert \mathbf{S_D} \rVert \) of the resulting vector can then be easily calculated. This value simultaneously represents the distance on the Poincaré sphere that is denoted as \( dS \). It therefore holds that:
\begin{equation}
dS=\lVert \mathbf{S_D} \rVert=\sqrt{\langle \mathbf{S_\text{D}},\mathbf{S_\text{D}} \rangle}
\label{formula:dS}
\end{equation}

If both vectors are identical, then the norm is zero. If they are mutually orthogonal (completely different), then the norm is equal to 2. This is because the norm represents the distance of two points on the Poincaré sphere. Since the vectors are normalized, the radius of this sphere is 1. The diameter is then equal to 2. In the present manuscript, \( dS \) is used only as a metric for evaluating the accuracy of the reconstructed polarization state. It is not used directly as a feedback variable for polarization control.

\section{Results}
\label{sectionResults}
Among other goals, the objective of this setup is to verify the accuracy of the LC polarimeter and to estimate the influence that the given inaccuracies may have on the resulting error rate of the quantum key distribution. Thus, two independent measurements and one simulation based on them are presented. These are as follows:

\begin{itemize}
    \item Measurement of the effect of the number of sub-measurements on accuracy.
    \item Measurement of the effect of LC switching time on accuracy.
    \item Simulation of the effect of the overall polarimeter accuracy on the QBER.
\end{itemize}
 To illustrate the effect of accuracy on the resulting QBER of the system, it is possible to consider QBER as a function of the norm \( dS \). 

\subsection{Dependence of accuracy on the number of measurements}
Both analytical methods described above require several partial intensity measurements. However, with the number of these, both the accuracy and the time increase. For this reason, it is necessary to determine their optimal number. This is performed by fully automated measurement in 256 iterations for 16 QWP and 16 HWP settings in the range 0° to 160°. First for the direct analysis with 4 values and then for Fourier analysis with 32, 16 and 8 values. In this way, every second value was always excluded. In this fashion, the Stokes vector was thus always determined from the same measurement, eliminating further inaccuracies.

The results can be found in the plots in Figure~\ref{Measurement1} below. From the data above the heatmaps, the average norm can be clearly seen, which in all cases is around 0.1. As expected, the norm increases (accuracy decreases) as the number of measurements decreases. However, although the measurement with 4 values has the lowest precision, the difference from the other three measurements is minimal. Hence, it can be chosen, especially due to the significant time savings, as it is two, four, and eight times faster than Fourier analysis.

\begin{figure}[h]
    \centering
    \includegraphics[width=0.8\linewidth]{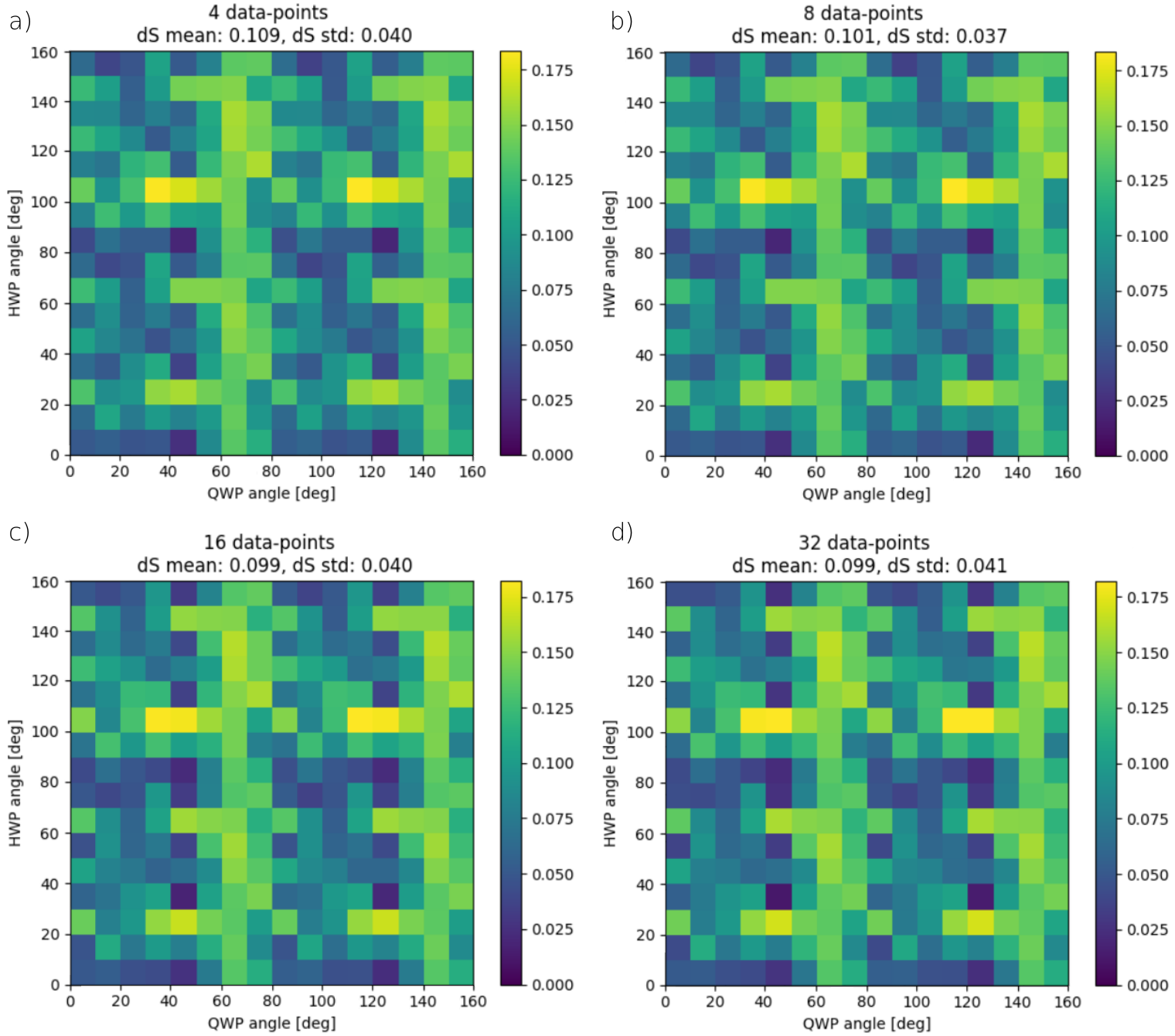}
    \caption{Accuracy of the determined polarization state for 256 generated polarization states and four different numbers of intensity sub-measurements: a) 4, b) 8, c) 16, and d) 32. Each heatmap shows the Euclidean norm \( dS \) between the calculated and measured Stokes vectors for a given combination of QWP and HWP angles. Lower values (darker blue) correspond to better agreement and thus higher polarimetric accuracy. The average value of \( dS \) and its standard deviation for each measurement configuration are indicated above the corresponding panel.}
    \label{Measurement1}
\end{figure}

At the same time, the inaccuracies in the setup are clearly visible in the ``blue-green pattern''. In fact, in a very ideal case, the whole heatmap should be dark blue. This pattern indicates recurring inaccuracies in different waveplate configurations. Among others, these might be caused by uncertainties in LCs characterization, slight component wavelength deviation, spatial inhomogeneities in polarization distribution, background noise, and imperfect alignment of the whole setup. However, for demonstration purposes, these errors are negligible. 

\subsection{Dependence of accuracy on the switching speed of LCs}
The second measurement investigates how the LC switching time affects the accuracy with which the target retardation values are reached during the sequence of polarimetric measurement configurations. In other words, the relevant switching time is the time required for the LC elements in the polarimeter to settle between successive analysis settings used to reconstruct a single polarization state. If the available switching time is too short, the LC retarders may not fully reach their target values, which leads to increased polarimetric error. The following measurement for LC switching values of 50, 100 and 200~ms monitors the norm by switching between two states of polarization:

\vspace{7pt}
\begin{itemize}
    \item \textbf{State I:} $\theta_{HWP} = \theta_{QWP} = 0$
    \item \textbf{State II:} $\theta_{HWP} = \theta_{QWP} = \frac{\pi}{16}$
\end{itemize}
\vspace{7pt}

Here state I corresponds to a simple reference configuration producing horizontal linear polarization, while State II represents a distinct non-trivial elliptical polarization state. These two input states were chosen as representative and well-reproducible test configurations. The aim of this experiment was not to investigate the full dependence on the input polarization state, but to evaluate the effect of LC switching dynamics on polarimetric accuracy. Since the dominant error in this case arises from the finite response of the LC actuators, these two clearly different states provide a representative test of the influence of switching time on the measurement accuracy.

\begin{figure}[h]
    \centering
    \includegraphics[width=0.8\linewidth]{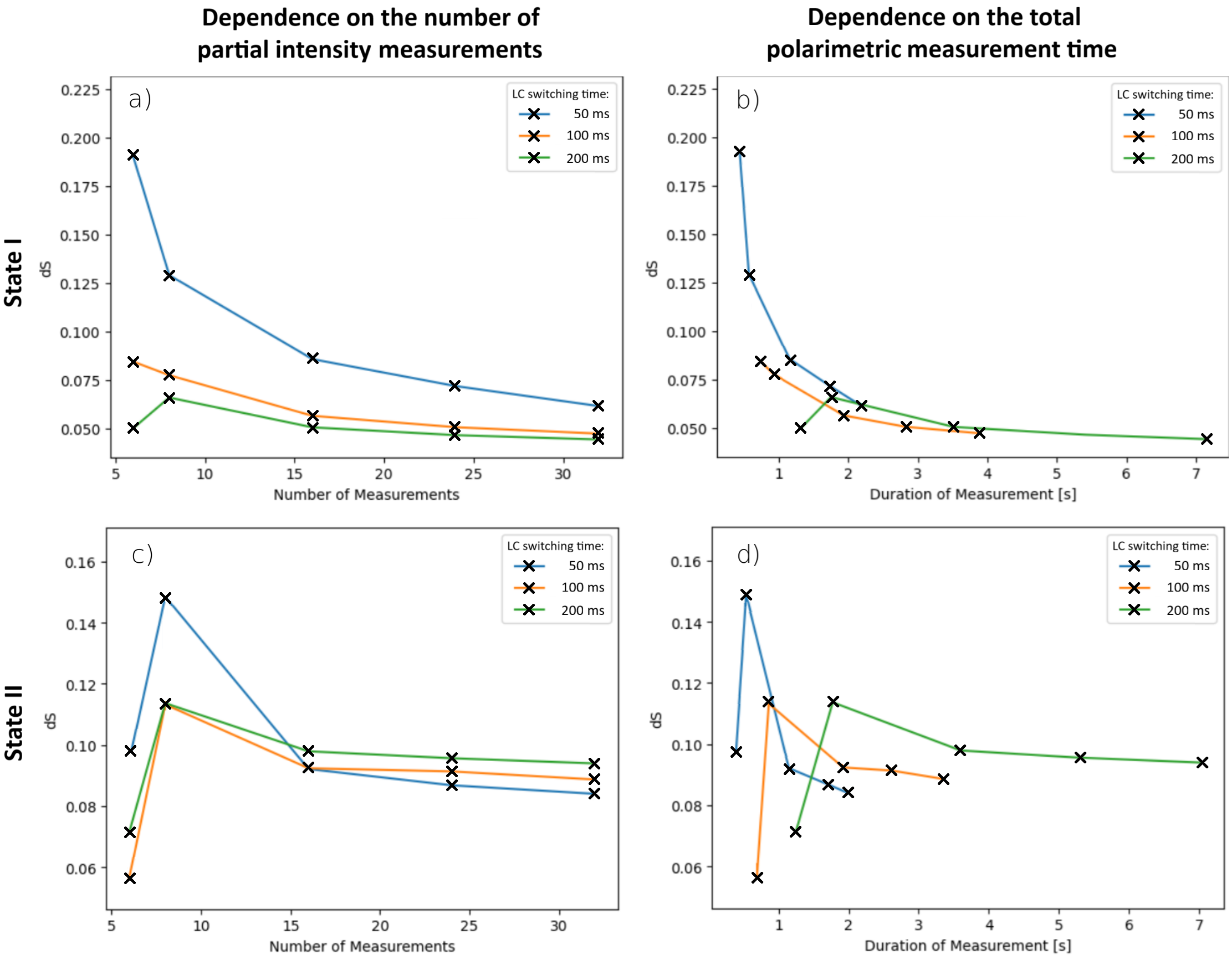}
    \caption{Impact of the LCs switching time on the accuracy of the total measurement, for different numbers of intensity sub-measurements (left) and the total polarimetric measurement time (right). For state I a), b) and for state II c), d).}
    \label{Measurement2}
\end{figure}

The plots on the left part of the Figure~\ref{Measurement2} show the accuracy of these measurements as a function of the number of partial intensity measurements (6, 8, 16, 24 and 32). These values were determined because a steeper function is expected at the beginning. The plots on the right were obtained by ``stretching'' the curves on the left, such that the x-axis reflects the total duration of the complete polarimetric measurement. This duration is determined, among other things, by both the LC switching time, the number of individual intensity measurements, and the constant exposure time, which was set to 20 ms.

In plots a) and c), for most of the function, the 50 ms switching time shows the worst results (especially at the beginning), while the results for both 100 and 200~ms are essentially comparable. However, as can be seen from plots b) and d), the total time for measuring the polarization state in the case of 100 ms is roughly half of that for 200~ms. Overall, it can be seen that the accuracy improves with longer measurement times, theoretically asymptotically to the limit of systematic uncertainty. For short measurement times, a~deterioration in accuracy can often be seen.


\subsection{Dependence of QBER on accuracy}

Based on the reconstructed polarization state of the reference laser, a corresponding set of compensation parameters can be determined for the polarization compensator and applied to the quantum channel. In the ideal case, this compensation fully corrects the polarization transformation introduced by the optical channel and preserves the expected correlations of the entangled state. In practice, the overall compensation accuracy may be limited by both the finite accuracy of the polarimetric measurement and by the finite performance of the polarization compensator itself. However, in the present analysis only the first one is considered and the compensation process is therefore assumed to be otherwise ideal. Under this assumption, any deviation of the implemented compensation parameters from their optimal values is attributed solely to the finite polarimetric accuracy, which leads to imperfect basis alignment and to an additional contribution to the quantum-bit error rate. In the following, this additional contribution is denoted as \(dQBER\).

To quantify this effect, a numerical simulation was performed for 10\,000 independent realizations of an entanglement-based E91 system. In this scheme, an entangled photon-pair source emits the Bell state:
\begin{equation}
\ket{\Psi}=\frac{\ket{HH}+\ket{VV}}{\sqrt{2}},
\label{eq:bell_state}
\end{equation}

Here, one photon propagates towards Alice and the other towards Bob through two separate optical paths. For each realization of the simulation, the two optical paths were represented by independent random unitary Jones matrices \(J_A\) and \(J_B\), describing the polarization transformations in the channels towards Alice and Bob. The liquid-crystal polarization compensator was described by the Jones matrix \(J_{{LC}}\). In the present simulation, the PCM is introduced only in one branch, corresponding to polarimetric reconstruction performed only on Alice's side of the link. The output state of the compensated system can therefore be written as:
\begin{equation}
\ket{\Psi_{\mathrm{out}}}=
\left(J_{{LC}} J_A \otimes J_B\right)\ket{\Psi}
\label{eq:output_state}
\end{equation}

This output state was then used to calculate the coincidence values and the corresponding additional contribution \(dQBER\). Since all other error sources were assumed to be zero in this simulation, this additional contribution is numerically equal to the total QBER. In addition, the Stokes-vector deviation \(dS\) was determined independently for the same realization. This made it possible to relate the polarimetric inaccuracy to the resulting increase in QBER. In the context of QKD, the coincidence values correspond to the simultaneous detection of entangled photons by Alice and Bob and are therefore associated with projections onto two-photon states \(\ket{ab}\), where \(a\) and \(b\) denote the polarization outcomes obtained by Alice and Bob, respectively, in the selected measurement basis. This yields the following correlated events: \(\ket{HH}\), \(\ket{VV}\), \(\ket{DD}\), and \(\ket{AA}\), as well as the error events \(\ket{HV}\), \(\ket{VH}\), \(\ket{DA}\), and \(\ket{AD}\). The coincidence probabilities \(C_{ab}\) were then obtained by projecting the transformed output state \(\ket{\Psi_{\mathrm{out}}}\) onto the corresponding two-photon measurement states \(\ket{ab}\):
\begin{equation}
C_{ab}=\left|\bra{ab}\Psi_{\mathrm{out}}\rangle\right|^2
\label{eq:coincidences}
\end{equation}

Here, the indices \(a\) and \(b\) correspond to the measurement results of Alice and Bob, respectively, with \(a,b \in \{H,V\}\) in the \(H/V\) basis and \(a,b \in \{D,A\}\) in the \(D/A\) basis. From the simulated coincidence values, the corresponding visibilities in the bases \(H/V\) and \(D/A\) can then be determined, from which the QBER for each basis is obtained separately.
\begin{equation}
\begin{split}
\mathrm{V}_{HV} &=
\frac{C_{HH}+C_{VV}-C_{HV}-C_{VH}}
{C_{HH}+C_{VV}+C_{HV}+C_{VH}}~~~~~~~~~~~\xrightarrow{}~~~~~~~~~~~QBER_{HV} &= \frac{1-\mathrm{V}_{HV}}{2}\\[8pt]
\mathrm{V}_{DA} &=
\frac{C_{DD}+C_{AA}-C_{DA}-C_{AD}}
{C_{DD}+C_{AA}+C_{DA}+C_{AD}}~~~~~~~~~~~~~\xrightarrow{}~~~~~~~~~~~QBER_{DA} &= \frac{1-\mathrm{V}_{DA}}{2}
\end{split}
\label{eq:visibilities_and_qber_bases}
\end{equation}

The overall QBER is obtained as the average of the values corresponding to the two measurement bases. Since all other contributions to the QBER are assumed to vanish in the present analysis, the additional contribution \(dQBER\) is equal to the total QBER:
\begin{equation}
dQBER=QBER=\frac{QBER_{HV}+QBER_{DA}}{2}
\label{eq:qber_total}
\end{equation}

In parallel, the polarimetric inaccuracy was quantified by comparing the ideal Stokes vector $\mathbf{S_\text{C}}$ of the perfectly compensated state with the Stokes vector $\mathbf{S_\text{M}}$ obtained for the perturbed controller settings. The corresponding error metric \(dS\) was defined as previously introduced in Equation~\ref{formula:dS}. Repeating this procedure for all simulated realizations yielded a set of corresponding \((dS,dQBER)\) values, which made it possible to statistically evaluate how the finite polarimetric accuracy affects the additional contribution to the QBER.

\begin{figure}[h]
    \centering
    \includegraphics[width=0.6\linewidth]{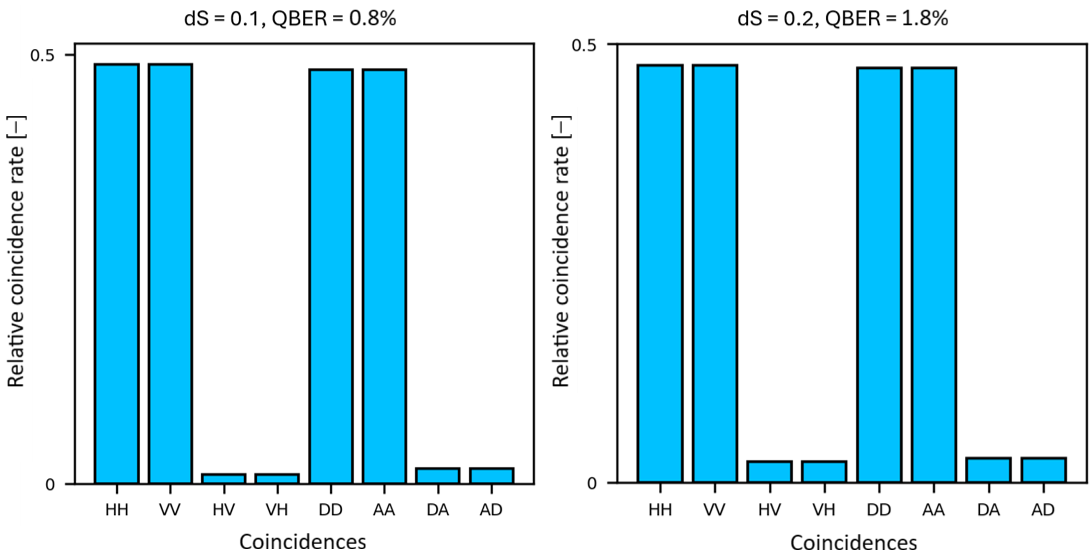}
    \caption{Two examples of relative coincidence distributions after polarization compensation for compensation settings affected by small errors.}
    \label{QBER1}
\end{figure}

Examples of the simulated coincidence distributions are shown in Figure~\ref{QBER1} for two representative cases with \(dS = 0.1\) and \(dS = 0.2\). These examples illustrate how an increasing mismatch of the compensated polarization state reduces the expected correlations and results in a larger additional contribution to the QBER. For these two representative cases, the corresponding values of \(dQBER\) are approximately \(0.8\,\%\) and \(1.8\,\%\), respectively.

\begin{figure}[h]
    \centering
    \includegraphics[width=0.4\linewidth]{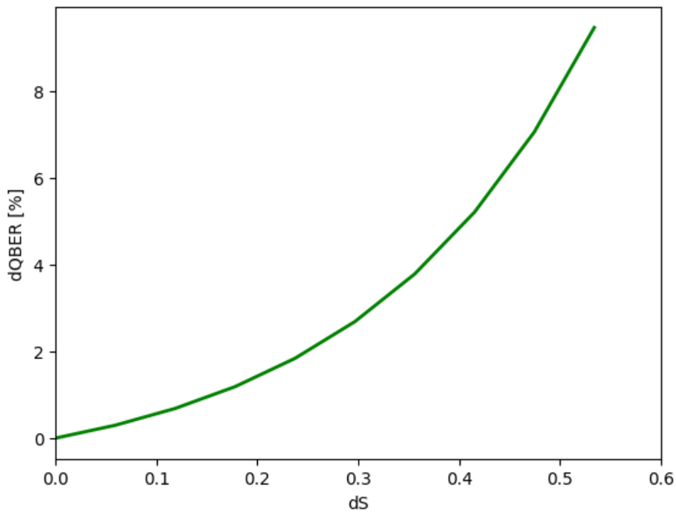}
    \caption{Increase of the quantum-bit error rate, denoted as \(dQBER\), as a function of the polarimetric inaccuracy \(dS\).}
    \label{QBER2}
\end{figure}

The overall dependence of \(dQBER\) on \(dS\), obtained from the full set of simulated realizations, is summarized in Figure~\ref{QBER2}. The results show that the additional QBER contribution caused by finite polarimetric accuracy increases with increasing \(dS\). Nevertheless, for the experimentally relevant range of \(dS\), the resulting increase remains sufficiently small such that secure key generation is still possible, although with a reduced key rate.

\newpage
\section{Conclusion}

This paper examined the use of liquid crystals for polarization monitoring and compensation in \textit{CubEniK} QKD system. The work focused on the design, implementation, and evaluation of a compact LC-based polarimeter. In particular, two separate experiments were carried out. The first compared Fourier-based analysis and direct analysis with respect to the number of measurements required for polarization-state reconstruction. The second examined the influence of LC switching time on the speed and accuracy of the reconstruction. In addition, a computer simulation was used to estimate how the experimentally observed reconstruction inaccuracy would translate into an additional contribution to the QBER. Overall, the results support LC-based polarimetry as a compact and fully electronic approach to polarization monitoring in future satellite QKD systems, with future work focusing on measurement optimization and integration into \textit{CubEniK} and related missions.

\section*{Acknowledgment}
The authors acknowledge support from a grant of the German Federal Ministry for Economic Affairs and Climate Action (BMWK) project number 50YH2205A (CubEniK). Furthermore, this work is supported by the Ministry of the Interior of the Czech Republic under grant \textcolor{black}{VJ03030019, project Expansion of international cooperation in the field of security of fiber optic networks, systems and services}.

\section*{Author contributions statement}
O.K., A.Z. and Y.F. conceived and performed the experiments and analyzed the data. O.V., P.M. and T.H. provided project administration, technical support and resources. All authors contributed to manuscript preparation and approved the final version.

\section*{Competing Interests}
The authors declare no competing interests.

\section*{Data Availability}
The data used and/or analyzed during the current study are not publicly available because they were prepared in collaboration with an industrial partner and are subject to confidentiality agreements. However, the data are available from the corresponding author upon reasonable request and with permission of the industrial partner.

\bibliography{sample}

\end{document}